\documentclass[prd,aps,preprint,tightenlines]{revtex4}
\usepackage{epsfig}
\usepackage{amsmath}
\vspace{1in}
\usepackage{axodraw}

\begin{document}
\title{Three-Quark Light-Cone Amplitudes of The Proton
  \\And Quark-Orbital-Motion Dependent Observables}
\author{Xiangdong Ji}
\email{xji@physics.umd.edu}
\affiliation{Department of Physics,
University of Maryland,
College Park, Maryland 20742, USA}
\author{Jian-Ping Ma}
\email{majp@itp.cn}
\affiliation{Institute of Theoretical Physics,
Academia Sinica, Beijing, 100080, P. R. China}
\author{Feng Yuan}
\email{fyuan@physics.umd.edu}
\affiliation{Department of Physics,
University of Maryland,
College Park, Maryland 20742, USA }

\date{\today}
\vspace{0.5in}          
\begin{abstract}

We study the three-quark light-cone amplitudes of the proton
including quarks' transverse momenta. We  
classify these amplitudes using a newly-developed method
in which light-cone wave functions
are constructed from a class of light-cone matrix elements.
We derive the constraints on the amplitudes from parity
and time-reversal symmetries. We use the amplitudes 
to calculate the physical observables which vanish when 
the quark orbital angular momentum is absent. These
include transverse-momentum dependent parton distributions
$\Delta q_T(x, k_\perp)$, $q_T(x, k_\perp)$, $\delta q(x, k_\perp)$, 
and $\delta q_L(x,k_\perp)$, twist-three
parton distributions $g_T(x)$ and $h_L(x)$, helicity-flip 
generalized parton distributions $E(x, \xi=0, Q^2)$ and its associates, 
and the Pauli form factor $F_2(Q^2)$. 

\end{abstract}

\maketitle
\def \p {\partial}
\def \dd {\phi_{u\bar dg}}
\def \ddp {\phi_{u\bar dgg}}
\def \pq {\phi_{u\bar d\bar uu}}
\def \jpsi {J/\phi}
\def \phip {\phi^\prime}
\def \to {\rightarrow}
\def \abst {\vert t \vert}
\def\bfsig{\mbox{\boldmath$\sigma$}}
\def\DT{\mbox{\boldmath$\Delta_T$}}
\def \jpsi {J/\phi}
\def\bfej{\mbox{\boldmath$\varepsilon$}}
\def\FT{T_R \vert_{t\to 0}}
\def\FV{T_V \vert_{t\to 0}}

\section{introduction}

In the simplest model of the proton, the valence quarks are moving in the 
$s$-orbit and the proton spin is constructed from the quark spins. 
This picture has been ruled out definitely by 
the EMC data \cite{Ashman:1989ig} and the follow-up experiments, SMC, E142, E143, E152, 
E153, and HERMES \cite{Filippone:2001ux}. One then expects that 
the quarks in the proton must have non-zero orbital angular momentum and/or the gluons 
must carry part of the proton spin. When the orbital angular momenta
of quarks and gluons are none-zero, the nucleon is intrinsically
deformed.  An interesting question is how big is the deformation? 
or how much does the quark orbital angular momentum contribute to the spin
of the nucleon? A few years ago one of us derived a sum rule, relating the
total angular momentum carried by quarks to an integral over the 
relevant generalized parton distributions (GPDs)\cite{Ji:1997ek,Muller:1994fv}.
The orbital angular momentum can be obtained by subtracting the measured
quark helicity contribution. This observation has lead to very 
active studies of GPDs and their measurements in hard exclusive 
processes\cite{Ji:1998pc,Radyushkin:2000uy,Goeke:2001tz}. 

There are, in fact, many observables which are potentially 
sensitive to the orbital angular momentum of the quarks, although 
they do not directly {\it measure} the orbital angular momentum itself. 
For example, the Pauli form factor $F_2(Q^2)$  of the proton, which 
has been measured recently using the recoil polarization technique at JLab 
\cite{Jones:1999rz,Gayou:2001qd}, has long been recognized as an
observable related to the light-cone amplitudes with non-zero 
orbital angular momentum \cite{Brodsky:1980zm,Jain:1993jf,Brodsky:2000ii}. 
The twist-three spin-dependent 
structure function $g_T(x)=g_1(x)+g_2(x)$ vanishes in the naive parton 
model, and its interpretation requires quarks' orbital 
motion\cite{Ioffe:1985ep,Kane:1978nd,Jaffe:1991qh}. 
The generalized parton distributions
such as $E(x, \xi, Q^2)$ were introduced to 
characterize quark orbital angular momentum
in the first place \cite{Ji:1997ek,Hoodbhoy:1998yb,Hoodbhoy:1998bt}. 
And finally, the transverse-momentum dependent 
parton distributions \cite{Soper:1979fq,Sivers:1990cc,
Sivers:1991fh,Anselmino:1995tv,Mulders:1996dh,Boer:1998nt} 
contain rich and direct information about the quark orbital 
motion \cite{Brodsky:2002cx,Brodsky:2002rv,Collins:2002kn,Ji:2002aa}. 
A natural question is can one correlate all these
observables in terms of light-cone wave functions in 
a minimally-model-dependent way? This paper is an attempt in 
this direction.

Our approach starts with the light-cone expansion of the proton
state truncated to the minimum Fock 
sector with just three valence quarks \cite{Lepage:1980fj,Brodsky:1998de}. 
Unlike the light-cone amplitudes studied in the exclusive processes
\cite{Chernyak:1984bm,King:1987wi,Chernyak:1989nv,
Braun:1999te,Braun:2000kw}, we keep
the full dependence on the transverse momenta of the quarks.
Some previous studies in this direction can found in 
Refs.\cite{Botts:1989kf,Bolz:1996sw,Diehl:1998kh}.
Clearly, this truncation is ideal: we do not
have the gluon and sea quark degrees of freedom; we do not have manifest
rotational symmetry and gauge invariance. However, the truncation
can be improved systematically by including higher Fock 
components, for example, adding one or more pair of sea quarks
and/or one or more gluons, etc. Therefore, in this study we will 
ignore certain artifacts arising from this specific truncation.

Based on our work here, one can go on to parameterize the three-quark
amplitudes and fit to the experimental data. The
amplitudes thus determined are phenomenological, 
and in principle cannot be compared with those 
solved from QCD directly. However, a comparison helps to determine 
the importance of the three quark amplitudes vs. 
multi-particle Fock amplitudes. One may also consider the 
phenomenological amplitudes as the effective proton wave function
after integrating out the gluon and sea quark degrees of freedom,
although the integrations are hard to implement in practice.

By committing ourselves to the light-cone amplitudes, 
we are also committing to the light-cone gauges $A^+=0$
\cite{Kogut:1970xa}. One can in principle add light-cone links 
between the quark fields to make the light-cone amplitudes 
gauge-invariant, but we do not find a compelling reason to do 
this. One subtlety about the light-cone gauge is that it requires 
additional gauge fixing \cite{Mueller:1986wy,Slavnov:1987yh,Antonov:1988yg}. 
Depending
on whether the additional gauge condition is time-reversal
invariant or not, the wave function amplitudes are real or fully complex. 
In the latter case, the final state interaction effects
are contained in the amplitudes \cite{Kovchegov:1997pc,Ji:2002aa,Belitsky:2002sm}.

Our plan of the presentation is as follows. We start in Sec. II by writing 
down the matrix elements of three-quark light-cone operators 
between proton state and the QCD vacuum, which serve to 
define a complete set of light-cone amplitudes within the 
truncation. We simplify these matrix elements
using color, flavor, spin, and discrete symmetries. 
We then invert them in Sec. III to find the three-quark light-cone 
wave functions. As we have explained in Ref. \cite{Burkardt:2002uc},
there are many advantages in this construction. One is
that the cut-off dependence of the light-cone amplitudes
can be studied directly from the matrix elements. We derived
the scale evolution equation of the individual Fock contributions to the parton
densities in \cite{Burkardt:2002uc}.  
Another is that these matrix elements may be calculated 
in the nucleon models or lattice QCD: it is not necessary to 
construct a model of the nucleon directly in the light-cone 
frame. Finally, the large-momentum behavior of the nucleon amplitudes
can be determined from Bethe-Salpeter-like equations governing
the matrix elements. In Sec. IV, we calculate
a number of physical observables which are directly linked 
to the amplitudes with non-zero quark orbital angular momentum. 

\section{Three-quark light-cone matrix elements}

In this section, we are interested in classifying the matrix elements
of three-quark operators between the proton and QCD vacuum. Since
much of the discussion depends on the use of light-cone coordinates
and light-cone gauge, we start by reminding the reader
some salient features of the light-cone technology, and
a more detailed exposition can be found in \cite{Lepage:1980fj,Brodsky:1998de}.
 
The light-cone time $x^+$ and coordinate $x^-$ are defined as $x^\pm
=1/\sqrt{2}(x^0\pm x^3)$. Likewise we define Dirac matrices
$\gamma^\pm = 1/\sqrt{2}(\gamma^0\pm\gamma^3)$. The 
projection operators for Dirac fields are defined 
as $P_\pm = (1/2)\gamma^\mp\gamma^\pm$. Any Dirac field $\psi$ 
can be decomposed into $\psi=\psi_++\psi_-$ with
$\psi_\pm = P_\pm \psi$. $\psi_+$ is a dynamical degrees of freedom 
and has the canonical expansion, 
\begin{eqnarray}
  \psi_+(\xi^+=0,\xi^-,\xi_\perp)
  &=& \int \frac{d^2k_\perp}{ (2\pi)^3}
    \frac{dk^+}{ 2k^+}
   \sum_\lambda
  \left[b_\lambda(k) u(k\lambda) e^{-i(k^+\xi^--\vec{k}_\perp\cdot\vec{\xi}_\perp)} \right. \nonumber \\
 && \left. + d_\lambda^\dagger(k) v(k\lambda)e^{i(k^+\xi^--\vec{k}_\perp\cdot\vec{\xi}_\perp)}
 \right]
  \ . 
\end{eqnarray}
Likewise, for the gluon fields in the light-cone gauge $A^+=0$, 
$A_\perp$ is dynamical and has the expansion,
\begin{eqnarray}
    A_\perp(\xi^+=0,\xi^-,\xi_\perp)
  &=&  \int \frac{d^2k_\perp}{(2\pi)^3}
    \frac{dk^+}{ 2k^+}
   \sum_\lambda
  \left[ a_\lambda(k) \epsilon(k\lambda) e^{-i(k^+\xi^--\vec{k}_\perp\cdot\vec{\xi}_\perp)}
   \right.
  \nonumber \\
  && \left. + a_\lambda^\dagger(k) \epsilon^*(k\lambda)e^{i(k^+\xi^--\vec{k}_\perp
   \cdot \vec{\xi}_\perp)} \right] \ . 
\end{eqnarray}
$\psi_-$ and $A^-$ are dependent variables.

The key observation in Ref. \cite{Burkardt:2002uc} 
is that the light-cone Fock expansion of a hadron state is 
{\it completely} defined by the 
matrix elements of a special class equal-light-cone-time 
quark-gluon operators between the QCD vacuum and the hadron. 
These operators are specified as follows: Take the + component
of the Dirac field $\psi_+$ and the $+\perp$-component of the
gauge field $F^{+\perp}$. [We sometimes label the $\perp$ components
with index $i=1, 2$.] Assume all these fields are at light-cone
time $x^+=0$, but otherwise with arbitrary dependence on other spacetime 
coordinates. Products of these fields with the right 
quantum numbers (spin, flavor, and color) define a set of  
operator basis. [This has been done in the
past for light-cone amplitudes in which all fields are separated along the
light-cone, see for example \cite{Braun:1999te}.] 
Clearly, these operators are not gauge-invariant 
although one can gauge-invariantize them by inserting light-cone links extending
from the locations of the fields to infinity for every field. 
Since all fields are at different spacetime points in the transverse 
direction, the operators do not require additional renormalization
subtractions apart from the usual wave function renormalization. 
The matrix elements of all these 
operators between the hadron state and the QCD vacuum contain {\it complete} 
information about the hadron wave function. 

In the following two subsections, we will use the above method to study
the matrix elements of operators constructed out of three quark fields
between proton and the QCD vacuum. We discuss separately the cases for 
the two up-quarks coupling to helicity-zero and one.

\subsection{Two Up-Quarks Coupling To Helicity-Zero}

When the two up-quarks are coupled to helicity-0, we define 
the following two matrix elements, 
\begin{eqnarray}
&& \langle 0|u^{Ta}_+(\xi_1)C\gamma^+ u_+^b(\xi_2) d_+^c(\xi_3)\epsilon^{abc} |P\rangle 
         \nonumber \\
  &=&  \phi^{(1)}(1,2,3) \gamma_5 U_+  
        + (i\partial^i_1\phi^{(2)}(1,2,3)
          + i\p_2^i \phi^{(2)}(2,1,3))   \gamma_5\gamma^{i}U_+  \nonumber \\
    &&      + \tilde \p_1^i \p_2^i \phi^{(3)}(1,2,3) U_+  
         \ ,    \nonumber  \\ ~~\\
&& \langle 0|u^{Ta}_+(\xi_1)C\gamma^+\gamma_5 u_+^b(\xi_2) d^c_+(\xi_3)
  \epsilon^{abc} |p\rangle  \nonumber \\
  &=&  \phi^{(4)}(1,2,3) U_+
        + (i\partial^i_1\phi^{(5)}(1,2,3)
        -i\partial^i_2\phi^{(5)}(2,1,3)) \gamma^{i} U_+ \nonumber \\
  &&      + \tilde \p_1^i \p_2^i \gamma_5 \phi^{(6)}(1,2,3) U_+  \ , 
\end{eqnarray}
where $a, b, c=(1,2,3)$ are color indices of the quarks, $C$ is the charge
conjugation matrix ($C=i\gamma^2\gamma^0$), and $|P\rangle$ 
is the proton state with momentum $P^\mu$ and Dirac spinor $U_+(P)$.
The arguments 1, 2, and 3 in the scalar functions $\phi^{(i)}$ 
stand for $(\xi_1^-,\vec{\xi}_{1\perp})$,
$(\xi_2^-,\vec{\xi}_{2\perp})$, and $\xi_3^-,\vec{\xi}_{3\perp})$, and
the functional dependence on transverse coordinates are of type 
$\vec{\xi}_{i\perp}\cdot \vec{\xi}_{j\perp}$ only.  
The up and down quark fields are represented by $u$ and $d$, 
respectively. The index $i$ goes over the transverse coordinates 1, 2. 
In the above equations, we have used the symmetry relations ($T$ stands for
transpose)
\begin{eqnarray}
          u^{Ta}_+(\xi_1)C\gamma^+ u_+^b(\xi_2)\epsilon^{abc}
            & = & u^{Ta}_+(\xi_2)C\gamma^+ u_+^b(\xi_1)\epsilon^{abc}  \ , \\ \nonumber 
          u^{Ta}_+(\xi_1)C\gamma^+\gamma_5 u_+^b(\xi_2)\epsilon^{abc} 
             & = & - u^{Ta}_+(\xi_2)C\gamma^+\gamma_5 u_+^b(\xi_1)\epsilon^{abc}  \ , 
\end{eqnarray}        
which imply 
\begin{eqnarray}
\phi^{(1)}(1,2,3) &=& \phi^{(1)}(2,1,3) \ ,   ~~~~ 
  \phi^{(6)}(1,2,3) = \phi^{(6)}(2,1,3)  \ ,
\nonumber\\
\phi^{(3)}(1,2,3) &=& -\phi^{(3)}(2,1,3) \ , ~~~~ 
\phi^{(4)}(1,2,3) = -\phi^{(4)}(2,1,3)  \ .
\end{eqnarray}
The matrix elements have an overall dependence on $\xi_1+\xi_2+\xi_3$
through a phase factor $e^{i(\xi_1+\xi_2+\xi_3)\cdot P/3}$, which we assumed
has been factorized {\it implicitly}. The remaining dependence is on 
the differences of the coordinates, for example, $\xi_1-\xi_2$
and $\xi_2-\xi_3$, because of the translational invariance.
Therefore, the partial derivative $\p_3^i$ is not independent 
of $\p_1^i$ and $\p_2^i$, and has been omitted. We have not included 
the structure $\tilde \partial^i\gamma^i$ in these equations because 
it is the same as $\partial^i\gamma^i\gamma_5$ when acting on $U_+$ 
using $\gamma^i\gamma^+\gamma^- U_+=0$. We use the shorthand
$\tilde \partial^i = \epsilon^{ij}\partial^j$ ($\epsilon^{12}=1$). Similarly the structure
$\tilde \p_1^i \p_2^i \partial^j_1 \gamma^{j}U_+$ can be reduced to the
existing ones. 

We can simplify the above equations by adding them
and applying the chiral projection operators $P_{L,R}
=(1\mp\gamma_5)/2$
\begin{eqnarray}
  && 2\langle 0|u^{Ta}_{+R}(\xi_1)C\gamma^+ u_{+L}^b(\xi_2) d_{+R}^c(\xi_3)
      \epsilon^{abc} |P\rangle 
         \nonumber  \\
  &=&  \left[\phi^{(1)}(1,2,3) -\phi^{(4)}(1,2,3)\right] U_{+R}
       + \tilde \p_1^i \p_2^i \left[\phi^{(3)}(1,2,3) -\phi^{(6)}(1,2,3)\right] U_{+R}
       \nonumber \\
       &&  + \left[i\partial^i_1\left(\phi^{(2)}(1,2,3) - 
         \phi^{(5)}(1,2,3)\right) +i\p_2^i\left(\phi^{(2)}(2,1,3))+\phi^{(5)}(2,1,3)\right)\right]  
            \gamma^{i}U_{+L} \ .
\end{eqnarray}
The two-up quarks have been paired to helicity-zero, and the remaining
down quark is projected to be right-handed. We can further simplify the above
equation by introducing the new amplitudes,
\begin{eqnarray}
  && \langle 0|u^{Ta}_{+R}(\xi_1)C\gamma^+ u_{+L}^b(\xi_2) d_{+R}^c(\xi_3)
      \epsilon^{abc}/\sqrt{6} |P\rangle 
         \nonumber  \\
  &=&  \psi^{(1)}(1,2,3) U_{+R} + i^2\p_1^i i\tilde\p_2^i
       \psi^{(2)}(1,2,3) U_{+R}  
        + \left[ i\partial^i_1 \psi^{(3)}(1,2,3) 
          +i\p_2^i \psi^{(4)}(1,2,3) \right]  \gamma^{i}U_{+L} \ ,
\end{eqnarray}
where $\psi$'s have no specific symmetry properties because the two 
up quarks are not in the same helicity state. If the nucleon
is right-handed, we split the above equation into two 
\begin{eqnarray}
 && \langle 0|u^{Ta}_{+R}(\xi_1)C\gamma^+ u_{+L}^b(\xi_2) d_{+R}^c(\xi_3)
      \frac{\epsilon^{abc}}{\sqrt{6}} |P \uparrow \rangle 
           = \left[\psi^{(1)}(1,2,3) + i^2 \p_1^i i\tilde \p_2^i
       \psi^{(2)}(1,2,3)\right] U_{+R}  \label{half1} \ , \\
&&  \langle 0|u^{Ta}_{+R}(\xi_1)C\gamma^+ u_{+L}^b(\xi_2) d_{+L}^c(\xi_3)
      \frac{\epsilon^{abc}}{\sqrt{6}} |P\uparrow \rangle 
  = \left[ i\partial^i_1 \psi^{(3)}(1,2,3) 
          +i\p_2^i \psi^{(4)}(1,2,3) \right]  \gamma^{i}U_{+R} \ .   
\label{nhalf1}
\end{eqnarray}
In our convention $\partial^i = -\partial/\partial x^i$ and $\gamma^i$
is that of Bjorken and Drell \cite{Bjorken:1979dk}. 

\subsection{Two Up-Quarks Coupling To Helicity-One}

When the two up quarks are coupled to helicity-1,
we define the matrix element,
\begin{eqnarray}
&& \langle 0|u^{aT}_{+}(\xi_1)Ci\sigma^{+i} u_{+}^b(\xi_2) d_{+}^c(\xi_3)
 \epsilon^{abc}/\sqrt{6} |P\rangle  \nonumber \\
   &=&  \phi^{(7)}(1,2,3) \gamma^{i} \gamma_5U_{+}
        + \left(i\partial^i_1 \phi^{(8)}(1,2,3)
       + i\partial^i_2 \phi^{(8)}(2,1,3) \right)\gamma_5  U_{+} \nonumber \\
        && +  \left(\tilde \partial^i_1 \phi^{(9)}(1,2,3)
       + \tilde \partial^i_2 \phi^{(9)}(2,1,3) \right)  U_{+} \nonumber \\
        && + \left[i\partial^{(i}_1 i\partial^{j)}_1 \phi^{(10)}(1,2,3)
      + i\partial^{(i}_2 i\partial^{j)}_2 \phi^{(10)}(2,1,3) \right.\nonumber \\
      && 
     \left.+ i\partial^{(i}_1 i\partial^{j)}_2  \phi^{(11)}(1,2,3)
       + i^2(\p_1^i\p_2^j-\p_2^i\p_1^j) \phi^{(12)}(1,2,3)\right]\gamma^{j} 
     \gamma_5 U_{+} \ ,  
\label{tensor1}
\end{eqnarray}
where the parentheses on a pair of indices indicate symmetrization 
and subtraction of the trace. The symmetry between two up-quarks yields
\begin{eqnarray}
\phi^{(7)}(1,2,3) &=& \phi^{(7)}(2,1,3) \ , ~~~~ 
\phi^{(11)}(1,2,3) = \phi^{(11)}(2,1,3) \ ,  \nonumber \\
\phi^{(12)}(1,2,3) &=& - \phi^{(12)}(2,1,3) \ . 
\end{eqnarray}
When inserting an additional $\gamma_5$ between the two up quark fields,
and using $\sigma^{+i}\gamma_5 = -i\epsilon^{ij}\sigma^{+j}$, 
we obtain a similar expression,
\begin{eqnarray}
&& \langle 0|u^{aT}_{+}(\xi_1)Ci\sigma^{+i}\gamma_5 u_{+}^b(\xi_2) d_{+}^c(\xi_3)
 \epsilon^{abc}/\sqrt{6} |P\rangle  \nonumber \\
   &=&  \phi^{(7)}(1,2,3)\gamma^{i} U_{+}
        + \left(\tilde \partial^i_1 \phi^{(8)}(1,2,3)
       + \tilde \partial^i_2 \phi^{(8)}(2,1,3) \right)\gamma_5  U_{+} \nonumber \\
        && +  \left( i\partial^i_1 \phi^{(9)}(1,2,3)
       +  i\partial^i_2 \phi^{(9)}(2,1,3) \right)  U_{+} \nonumber \\
        && - i\left [i\tilde \partial^{(i}_1i\partial^{j)}_1 \phi^{(10)}(1,2,3)
      + i\tilde \partial^{(i}_2 i \partial^{j)}_2 \phi^{(10)}(2,1,3)
        \right. \nonumber \\ && 
     +\left. i\tilde \partial^{(i}_1i \partial^{j)}_2  \phi^{(11)}(1,2,3)
       + i^2(\tilde \partial^i_1\partial^j_2-\tilde \partial^i_2\partial^j_1)
       \phi^{(12)}(1,2,3)\right]\gamma^{j} 
     \gamma_5 U_{+} \ , 
\label{tensor2} 
\end{eqnarray}
where we have used $\gamma^i=-i\epsilon^{ij}\gamma^j\gamma_5$,
valid when acting on $U_+$.

Combining Eqs. (\ref{tensor1}) and (\ref{tensor2}), we obtain 
the matrix elements of the quark fields in definite chirality states. 
For example, when the two up-quarks are right-handed, and the down quark
left-handed, the total helicity of the quarks is 1/2. In this case,
we have
\begin{equation}
  \langle 0|u^{Ta}_{+R}(\xi_1)Ci\sigma^{+i} u_{+R}^b(\xi_2) d_{+L}^c(\xi_3)
      \epsilon^{abc}/\sqrt{6} |P \uparrow \rangle 
       =   \left[\phi^{(7)}(1,2,3)
       + i(i\partial_1i\tilde \partial_2)\phi^{(12)}(1,2,3)\right] \gamma^i U_{+R} \ ,
\end{equation}
where we have used the relations $\gamma^i U_{+R} = -i\tilde \gamma^i U_{+R}$ 
and $\partial^{(i-}\partial^{j)}\gamma^jU_{R+}=0$ with $\p_1^{i-} = 
\p_1^i - i\epsilon^{ij} \p_1^j$.  
The isospin symmetry imposes the following constraint, 
\begin{eqnarray}
 && \langle 0|u^{Ta}_{+R}(\xi_1)Ci\sigma^{+i} u_{+R}^b(\xi_2) d_{+L}^c(\xi_3)
      \epsilon^{abc} |P \uparrow \rangle \nonumber \\
+ && \langle 0|d^{Ta}_{+R}(\xi_1)Ci\sigma^{+i} u_{+R}^b(\xi_2) u_{+L}^c(\xi_3)
      \epsilon^{abc} |P \uparrow\rangle \nonumber \\
+       && \langle 0|u^{Ta}_{+R}(\xi_1)Ci\sigma^{+i} d_{+R}^b(\xi_2) u_{+L}^c(\xi_3)
      \epsilon^{abc} |P \uparrow \rangle =0 \ . 
\end{eqnarray}
Using Fierz identity, one can show
\begin{eqnarray}
u^{Ta}_{+R}(\xi_1)Ci\sigma^{+i} d_{+R}^b(\xi_2) u_{+L}^c(\xi_3)
      \epsilon^{abc} 
 &=& - u^{Ta}_{+R}(\xi_1)C \gamma^+ d_{+R}^b(\xi_2) \gamma^i u_{+L}^c(\xi_3)
      \epsilon^{abc}\ ,   \nonumber \\
d^{Ta}_{+R}(\xi_1)Ci\sigma^{+i} u_{+R}^b(\xi_2) u_{+L}^c(\xi_3)
      \epsilon^{abc} 
 &=& - u^{Ta}_{+R}(\xi_2)C \gamma^+ d_{+R}^b(\xi_3) \gamma^i u_{+L}^c(\xi_1)
      \epsilon^{abc}\ , 
\end{eqnarray}
which means that $\phi^{(7,12)}$ are not independent amplitudes, 
\begin{eqnarray}
     \phi^{(7)}(1,2,3) &=& \psi^{(1)}(1,3,2) + \psi^{(1)}(2,3,1)   \ ,  \nonumber \\
     \phi^{(12)}(1,2,3) &=& \psi^{(2)}(1,3,2) - \psi^{(2)}(2,3,1)  \ .
\end{eqnarray}
Hence we write, 
\begin{eqnarray}
  && \langle 0|u^{Ta}_{+R}(\xi_1)Ci\sigma^{+i} u_{+R}^b(\xi_2) d_{+L}^c(\xi_3)
      \frac{\epsilon^{abc}}{\sqrt{6}} |P \uparrow \rangle   \nonumber \\
       &=&   \left[\psi^{(1)}(1,3,2)+\psi^{(1)}(2,3,1)   
       + i(i\partial_1i\tilde \partial_2)\left(\psi^{(2)}(1,3,2) -\psi^{(2)}(2,3,1) 
         \right)\right] \gamma^i U_{+R} \ , 
\label{half2}
\end{eqnarray}
which contains no new amplitude.

If all quarks are right-handed, the total helicity of the quarks
is 3/2. We need one unit of orbital angular momentum (projection on the
z-direction) to construct
the helicity of the proton. We write, 
\begin{eqnarray}
 && 2\langle 0|u^{Ta}_{+R}(\xi_1)Ci\sigma^{+i} u_{+R}^b(\xi_2) d_{+R}^c(\xi_3)
     \epsilon^{abc}/\sqrt{6} |P \uparrow \rangle \nonumber \\
          & =& \left[i\p_1^{i-}\phi^{(8+9)}(1,2,3) 
                   + i\p_2^{i-}\phi^{(8+9)}(2,1,3)\right]U_{+R}\ , 
\end{eqnarray}
where $\phi^{(8+9)}=\phi^{(8)}+\phi^{(9)}$. 
The isospin symmetry leads to the following relation 
\begin{equation}
  \phi^{(8+9)}(1,2,3)+\phi^{(8+9)}(1,3,2)-\phi^{(8+9)}(3,1,2)-\phi^{(8+9)}(3,2,1)=0 \ . 
\end{equation}
If we define 
\begin{equation} 
        \psi^{(8+9)}(1,2,3)/2 = \psi^{(5)}(1,2,3) - \psi^{(5)}(1,3,2) \ , 
\end{equation}
the above constraint is solved. Therefore, 
\begin{eqnarray}
 && \langle 0|u^{Ta}_{+R}(\xi_1)Ci\sigma^{+i} u_{+R}^b(\xi_2) d_{+R}^c(\xi_3)
     \frac{\epsilon^{abc}}{\sqrt{6}} |P \uparrow \rangle \nonumber \\
          & =& \left[i\p_1^{i-}\left(\psi^{(5)}(1,2,3) - \psi^{(5)}(1,3,2)\right) 
                   + i\p_2^{i-}\left(\psi^{(5)}(2,1,3) - \psi^{(5)}(2,3,1)\right) 
\right]U_{+R}\ , 
\label{threehalf}
\end{eqnarray}
which defines a new amplitude $\psi^{(5)}$. 

When the two up quarks are left-handed and the down quark right-handed,
the total quark helicity is $-1/2$. We again need one unit of orbital 
angular momentum to construct the proton helicity. The orbital angular momentum
can either be on the first or second particle, 
\begin{eqnarray}
 && 2\langle 0|u^{Ta}_{+L}(\xi_1)Ci\sigma^{+i} u_{+L}^b(\xi_2) d_{+R}^c(\xi_3)
     \epsilon^{abc}/\sqrt{6} |P \uparrow \rangle \nonumber \\
          & =&  \left[i\p_1^{i+} \phi^{(8-9)}(1,2,3) + i\p_2^{i+} \phi^{(8-9)}(2,1,3)
             \right]U_{+R} \ ,  
\end{eqnarray}
where $\p_1^{i+} = \p_1^i + i\epsilon^{ij} \p_1^j$ and so on.
Using the constraint from the isospin symmetry, one can express
$\phi^{(8-9)}$ in terms of $\psi^{(3)}$ and $\psi^{(4)}$, 
\begin{equation}
    \phi^{(8-9)} (1,2,3)/2 = \psi^{(4)}(3,1,2) - \psi^{(3)}(3,1,2)-\psi^{(3)}(3,2,1) \ . 
\end{equation}
Hence we write
\begin{eqnarray}
 && \langle 0|u^{Ta}_{+L}(\xi_1)Ci\sigma^{+i} u_{+L}^b(\xi_2) d_{+R}^c(\xi_3)
     \frac{\epsilon^{abc}}{\sqrt{6}} |P \uparrow \rangle \nonumber \\
          & =&  \left[i\p_1^{i+} \left(\psi^{(4)}(3,1,2) - \psi^{(3)}(3,1,2)
          - \psi^{(3)}(3,2,1)\right) 
           + (1\leftrightarrow 2) 
             \right]U_{+R} \ ,  
\label{nhalf2}
\end{eqnarray}
which does not contain any new amplitude.

When all quarks are left-handed, the total quark helicity is $-3/2$; 
two units of orbital angular momentum are need to construct the 
nucleon helicity. Taking the appropriate projection of the above equations, 
we get
\begin{eqnarray}
 && 2\langle 0|u^{Ta}_{+L}(\xi_1)Ci\sigma^{+i} u_{+L}^b(\xi_2) d_{+L}^c(\xi_3)
      \epsilon^{abc}/\sqrt{6} |P \uparrow \rangle \nonumber \\
  & = & \left[i\p_1^{(i+} i\p_1^{j)}\phi^{(10)}(1,2,3) 
          + i\p_2^{(i+} i\p_2^{j)}\phi^{(10)}(2,1,3)
         +  i\p_1^{(i+} i\p_2^{j)}\phi^{(11)}(1,2,3)\right]
          \gamma^j U_{+R} \ , 
\label{dear}
\end{eqnarray}
where $\phi^{(10)}$ describes the state in which the same quark carrys two units
of angular momentum, and $\phi^{(11)}$ describes each up quark 
carrying one unit of angular 
momentum. The constraint from the isospin symmetry 
relates $\phi^{(11)}$ to $\phi^{(10)}$
\begin{equation}
      \phi^{(11)}(1,2,3)
         = \phi^{(10)}(1,2,3) + \phi^{(10)}(2,1,3)
        -\phi^{(10)}(1,3,2) - \phi^{(10)}(2,3,1) \ . 
\end{equation}
Following our convention, we rename $\phi^{(10)}(1,2,3)/2$ to $\psi^{(6)}(1,2,3)$, 
and Eq. (\ref{dear}) becomes 
\begin{eqnarray}
 && \langle 0|u^{Ta}_{+L}(\xi_1)Ci\sigma^{+i} u_{+L}^b(\xi_2) d_{+L}^c(\xi_3)
      \frac{\epsilon^{abc}}{\sqrt{6}} |P \uparrow \rangle \nonumber \\
  & = & \left[i\p_1^{(i+} i\p_1^{j)}\psi^{(6)}(1,2,3) 
          + i\p_2^{(i+} i\p_2^{j)}\psi^{(6)}(2,1,3)\right. \nonumber \\
       && \left. + i\p_1^{(i+} i\p_2^{j)}\left(\psi^{(6)}(1,2,3)
        + \psi^{(6)}(2,1,3) - \psi^{(6)}(1,3,2) - \psi^{(6)}(2,3,1)\right)\right]
          \gamma^j U_{+R} \ ,
\label{nthreehalf}
\end{eqnarray}
which contains the new amplitude $\psi^{(6)}$. 

\section{The three-quark light-cone wave function of the proton}

The matrix elements considered in the previous section can be calculated
in models such as the MIT bag or QCD sum rules or lattice field theory. 
Once they are known, one can invert the equations to get the
light-cone wave function of the proton in the minimal Fock sector.
In this section, we work out the light-cone wave function
in terms of these matrix elements. The expression allows us to calculate
the physical observables of the proton in the next section.

Consider first the case in which all quarks are coupled to helicity 1/2.
Define the measure for the quark momentum integrations,
\begin{eqnarray}
    d[1]d[2]d[3] &=& \sqrt{2}\frac{dx_1dx_2dx_3}{\sqrt{2x_1 2x_2 2x_3}}
                  \frac{d^2\vec{k}_{1\perp}d^2
             \vec{k}_{2\perp}d^2\vec{k}_{3\perp}}{(2\pi)^9} 
       \nonumber \\
                  && \times 2\pi\delta(1-x_1-x_2-x_3)(2\pi)^2\delta^{(2)}
                   (\vec{k}_{1\perp}+\vec{k}_{2\perp}+\vec{k}_{3\perp}) \ .
\end{eqnarray}
where $x_i$ are the fraction of the nucleon momentum carried
by the quarks, and $\vec{k}_{i\perp}$ are their transverse momenta.
We introduce a Fock component for a right-handed proton, 
\begin{eqnarray}
  |P\uparrow\rangle_{1/2} &=& \int d[1]d[2]d[3]\left(\tilde \psi^{(1)}(1,2,3)
         + i(k_1^xk_2^y-k_1^yk_2^x) \tilde \psi^{(2)}(1,2,3)\right) \nonumber \\
         &&  \times  \frac{\epsilon^{abc}}{\sqrt{6}} u^{\dagger}_{a\uparrow}(1)
             \left(u^{\dagger}_{b\downarrow}(2)d^{\dagger}_{c\uparrow}(3)
            -d^{\dagger}_{b\downarrow}(2)u^{\dagger}_{c\uparrow}(3)\right)
         |0\rangle \ , 
\end{eqnarray}
where $\tilde \psi^{(1,2)}$ are functions of quark momenta with argument 1
represents $x_1$ and $k_{1\perp}$ and so on. 
The dependence on the transverse momenta is of form 
$\vec{k}_{i\perp}\cdot\vec{k}_{j\perp}$ only. The $\tilde \psi^{(2)}$
amplitude has a pre-factor $k_1^xk_2^y-k_1^yk_2^x$ which signals the 
contributions from quarks with non-zero orbital angular momentum, 
although all magnetic quantum numbers sum to zero. 
Using the above state, the matrix elements shown in 
Eqs. (\ref{half1}) and (\ref{half2}) can be calculated. 
We find the relation between the 
wave function amplitudes and the matrix elements 
\begin{eqnarray}
  \psi^{(\alpha)}(\xi_1,\xi_2,\xi_3)
     &=& \int d[1]d[2]d[3] \sqrt{x_12x_22x_3}
                 ~ e^{-i(x_1P^+\xi_1^--\vec{k}_{1\perp}\cdot \vec{\xi}_{1\perp})}
 \nonumber \\
&& \times e^{-i(x_2P^+\xi_2^--\vec{k}_{2\perp}\cdot \vec{\xi}_{2\perp})}
 e^{-i(x_3P^+\xi_3^--\vec{k}_{3\perp}\cdot \vec{\xi}_{3\perp})}
  \tilde \psi^{(\alpha)}(1,2,3) \ , 
\label{fourier}
\end{eqnarray}
where $\alpha=1,2$. 

Turn to the case where the quark helicities sum to $-1/2$.  
The corresponding Fock component can be written as
\begin{eqnarray}
  |P\uparrow\rangle_{-1/2} &=& \int d[1]d[2]d[3]\left((k_1^x+ik_1^y)
          \tilde \psi^{(3)}(1,2,3)
         + (k_2^x+ik_2^y) \tilde \psi^{(4)}(1,2,3)\right) \nonumber \\
         &&  \times  \frac{\epsilon^{abc}}{\sqrt{6}} \left( u^{\dagger}_{a\uparrow}(1)
            u^{\dagger}_{b\downarrow}(2)d^{\dagger}_{c\downarrow}(3)
            -d^{\dagger}_{a\uparrow}(1)u^{\dagger}_{b\downarrow}(2)
             u^{\dagger}_{c\downarrow}(3)\right)
         |0\rangle \ . 
\end{eqnarray}
The matrix elements in Eqs. (\ref{nhalf1}) and (\ref{nhalf2}) can be
calculated, and the resulting $\psi^{(3,4)}$ are related to the above 
amplitudes in the same way as in Eq. (\ref{fourier}). 
One might suspect if additional independent amplitudes can 
be constructed by adding terms with structure
$k_1^xk_2^y-k_1^yk_2^x$. A careful examination indicates 
that they can be reduced to the already existing ones. 

When the quark helicity is added to $3/2$, the Fock component 
is 
\begin{eqnarray}
  |P\uparrow\rangle_{3/2} &=& \int d[1]d[2]d[3]~(-k_1^x+ik_1^y)
          \tilde \psi^{(5)}(1,2,3) \nonumber \\
         &&  \times  \frac{\epsilon^{abc}}{\sqrt{6}} u^{\dagger}_{a\uparrow}(1)
             \left(
     u^{\dagger}_{b\uparrow}(2)d^{\dagger}_{c\uparrow}(3)
    -d^{\dagger}_{b\uparrow}(2)u^{\dagger}_{c\uparrow}(3)
            \right)
         |0\rangle \ .
\end{eqnarray}
Calculating the matrix element in Eq. (\ref{threehalf}), we find
$\psi^{(5)}(\xi_1,\xi_2,\xi_3)$ is the Fourier transformation of $\tilde\psi^{(5)}
(k_1,k_2,k_3)$. 

Finally, we consider the case when the quark helicity adds to $-3/2$, 
the Fock component is
\begin{eqnarray}
  |P\uparrow\rangle_{-3/2} &=& \int d[1]d[2]d[3]~(k_1^x+ik_1^y)(k_3^x+ik_3^y)
         \tilde \psi^{(6)}(1,2,3) \nonumber \\
         &&  \times  \frac{\epsilon^{abc}}{\sqrt{6}} u^{\dagger}_{a\downarrow}(1)
             \left(d^{\dagger}_{b\downarrow}(2)u^{\dagger}_{c\downarrow}(3)-u^{\dagger}_{b\downarrow}(2)d^
{\dagger}_{c\downarrow}(3)
            \right)
         |0\rangle \ .
\end{eqnarray}
Using this, we calculate the matrix elements in Eq. (\ref{nthreehalf})
and find $\psi^{(6)}(\xi_1,\xi_2,\xi_3)$ is just the Fourier transformation 
of $\tilde \psi^{(6)}(k_1,k_2,k_3)$. 

The complete three-quark light-cone Fock expansion of the proton  
has the following form, 
\begin{equation}
   |P\uparrow \rangle = |P\uparrow\rangle_{-3/2} 
         + |P\uparrow\rangle_{-1/2} 
          + |P\uparrow\rangle_{1/2} 
         + |P\uparrow\rangle_{3/2} \ .  
\end{equation}
Many interesting proton observables can be calculated using the above wave function
as we will show in the next section.

Using the results in the previous section, one can also 
construct a proton state with the negative helicity $|P\downarrow\rangle$. 
All the wave function amplitudes are the same, except that the 
quark helicities are flipped,  $k^x\pm ik^y$ becomes $k^x\mp ik^y$, 
and some signs are added,
\begin{eqnarray}
  |P\downarrow\rangle_{-1/2} &=& \int d[1]d[2]d[3]\left(-\tilde \psi^{(1)}(1,2,3)
         + i(k_1^xk_2^y-k_1^yk_2^x) \tilde \psi^{(2)}(1,2,3)\right) \nonumber \\
         &&  \times  \frac{\epsilon^{abc}}{\sqrt{6}} u^{\dagger}_{a\downarrow}(1)
             \left(u^{\dagger}_{b\uparrow}(2)d^{\dagger}_{c\downarrow}(3)
            -d^{\dagger}_{b\uparrow}(2)u^{\dagger}_{c\downarrow}(3)\right)
         |0\rangle \ , 
\end{eqnarray}
\begin{eqnarray}
  |P\downarrow\rangle_{1/2} &=& \int d[1]d[2]d[3]\left((k_1^x-ik_1^y)
          \tilde \psi^{(3)}(1,2,3)
         + (k_2^x-ik_2^y) \tilde \psi^{(4)}(1,2,3)\right) \nonumber \\
         &&  \times  \frac{\epsilon^{abc}}{\sqrt{6}} \left( u^{\dagger}_{a\downarrow}(1)
            u^{\dagger}_{b\uparrow}(2)d^{\dagger}_{c\uparrow}(3)
            -d^{\dagger}_{a\downarrow}(1)u^{\dagger}_{b\uparrow}(2)
             u^{\dagger}_{c\uparrow}(3)\right)
         |0\rangle \ , 
\end{eqnarray}
\begin{eqnarray}
  |P\downarrow\rangle_{-3/2} &=& \int d[1]d[2]d[3]~(-)(k_1^x+ik_1^y)
          \tilde \psi^{(5)}(1,2,3) \nonumber \\
         &&  \times  \frac{\epsilon^{abc}}{\sqrt{6}} u^{\dagger}_{a\downarrow}(1)
             \left(
     u^{\dagger}_{b\downarrow}(2)d^{\dagger}_{c\downarrow}(3)
          -d^{\dagger}_{b\downarrow}(2)u^{\dagger}_{c\downarrow}(3)  \right)
         |0\rangle \ , 
\end{eqnarray}
\begin{eqnarray}
  |P\downarrow\rangle_{3/2} &=& \int d[1]d[2]d[3]~(-)(k_1^x-ik_1^y)(k_3^x-ik_3^y)
         \tilde \psi^{(6)}(1,2,3) \nonumber \\
         &&  \times  \frac{\epsilon^{abc}}{\sqrt{6}} u^{\dagger}_{a\uparrow}(1)
             \left(d^{\dagger}_{b\uparrow}(2)u^{\dagger}_{c\uparrow}(3)-
            u^{\dagger}_{b\uparrow}(2)d^{\dagger}_{c\uparrow}(3)      \right)
         |0\rangle \  . 
\end{eqnarray}
The same expressions can be obtained from Jacob and Wick's method \cite{Jacob:1959at},
\begin{equation}
                     (-1)^{s-\lambda}|P-\lambda\rangle = \hat Y|P\lambda\rangle \ , 
\end{equation}
where $\hat Y$ is the parity operation followed by a $180^\circ$ rotation 
in the $y$-direction, and $s$ is the spin and $\lambda$ the helicity.

\subsection{Constraints From Time-Reversal And Parity}

Consider the proton state $|P\uparrow\rangle$. Under parity, the 3-components
of the proton momentum change the sign: $(P^0, \vec{P})$ becomes
$(P^0, -\vec{P})$. Under time-reversal, the momentum changes in the same
way. Therefore under combined time-reversal and parity, the momentum of the 
proton does not change. Neither do the quark momenta. On the other hand, 
helicity changes sign 
under the combined parity and time reversal, and hence the proton state 
$|P\uparrow\rangle$ becomes $|P\downarrow\rangle$. Similar changes
occur for the quark states.

Under time-reversal, one must replace C-numbers with their complex conjugates.
Thus under the combined time reversal and parity, a positive-helicity 
proton state becomes a negative helicity one $|P\downarrow\rangle$ 
with wave function amplitudes complex-conjugated, quark helicity flipped, 
and $k^x\pm ik^y$ becoming $k^x\mp ik^y$. 
Moreover, there is a sign change of $(-1)^{s-\lambda}$ from time-reversal. 
The resulting state $|P\downarrow\rangle$ is exactly the same 
as that in Eqs. (36-39), except
all amplitudes are complex conjugated. Hence if time-reversal symmetry applies,
all the wave function amplitudes must be real, 
\begin{equation}
          \tilde \psi^{(i)*} = \tilde \psi^{(i)},~~~~ i = 1, 2,3,4,5,6 \ ,   
\end{equation}
which is true in many model calculations.

The above result, however, is only correct if the gauge condition 
is also invariant under the discrete symmetries. 
In light-cone gauge, $A^+=0$ does 
not fix the gauge completely, additional boundary conditions for 
the gauge field must be specified \cite{Mueller:1986wy,Slavnov:1987yh,Antonov:1988yg}. 
This additional gauge fixing might
not be invariant under the combined parity and time-reversal.
For example, the advanced and retarded boundary conditions transform
into each other under time-reversal; however the principal 
value prescription is self-conjugating.
In the former case, the wave function amplitudes 
are no longer constrained by the above condition. 
Therefore, although the final-state-interaction gauge links 
vanish in the momentum-dependent parton distributions in the light-cone gauge, 
the imaginary part of the wave function amplitudes
can reproduce the final state interaction 
effects \cite{Brodsky:2002cx, Brodsky:2002rv, Collins:2002kn,Ji:2002aa}. 
In the following discussion, 
we use the advanced boundary condition and 
hence the imaginary parts of the amplitudes do not vanish.

\section{Observables depending on Quark  
orbital angular momentum}
 
As we have explained in Introduction, the quark orbital 
angular momentum in the proton is certainly non-zero. To determine
it experimentally, one has to measure the relevant generalized
parton distributions \cite{Ji:1997ek}. On the other hand, there are many other
nucleon observables which are sensitive to the quark
orbital angular momentum, although they do not measure 
it directly. The most familiar example
is the Pauli form factor $F_2(Q^2)$ of the nucleon.
In light-cone quantization, $F_2(Q^2)$ is an 
helicity-flip observable which depends on the 
interference between the wave function amplitudes differing
in one unit of orbital angular momentum \cite{Brodsky:1980zm,
Jain:1993jf}. 

We have to caution, however, that when the gauge fields are 
taken into account explicitly, some orbital angular momentum 
effects discussed here cannot be cleanly separated from the 
gluon contributions. In fact, the quark angular momentum 
\begin{equation}
           L_q =\int d^3x \psi^\dagger \vec{x}\times (-i\vec{D}) \psi \ , 
\end{equation}
does contain the gluon contribution through the gauge potential in the covariant 
derivative. In the example of the electron magnetic moment
in quantum electrodynamics, the photon Fock component plays an 
important role \cite{Brodsky:1980zm}. Hence, some results in this
section can only make sense if considered as an effective description
after integrating out the gluon degrees of freedom. 

In the quark models, the need for a quark
orbital angular momentum can sometimes be avoided by 
large constituent quark masses. The quarks can have 
significant spin flips through the mass term. Then some observables 
discussed in this section are directly proportional to the 
constituent masses \cite{Feynman:1972}. In QCD, because the light 
flavors have negligible masses, the quark spin flips cannot occur
and the orbital angular momentum and gluon angular momentum are 
thereby essential.

We start by calculating the hadron helicity-flip 
transverse-momentum dependent parton distributions. 
The twist-three distributions
are found to be simply the integrals over the appropriate 
transverse-momentum dependent distributions. After this, we 
calculate the generalized
parton distribution and electromagnetic form factors, with
$E(x, \xi, Q^2)$ and $F_2(Q^2)$ as examples.  

\subsection{Transverse-Momentum Dependent Parton Distributions}

The transverse momentum dependent parton distributions were
first introduced by Collins and Soper to describe Drell-Yan production
\cite{Soper:1979fq,Collins:1981uk} and later by Sivers to describe the single
spin asymmetry in hadron-hadron scattering
\cite{Sivers:1990cc,Sivers:1991fh}. A 
classification of the leading distributions in terms of spin
and chirality can be found in Ref. \cite{Mulders:1996dh,Boer:1998nt}. 
Recently, there has been much interest in measuring these 
distributions in the semi-inclusive deep-inelastic scattering
\cite{Airapetian:1999ib,Airapetian:2001yk,Brodsky:2002cx,
Anselmino:1997qx,Mulders:1996dh,Boer:1998nt,Efremov:2001cz,Ma:2001ie,Bacchetta:2002tk}.

Let us first recapitulate the classification, and in the process,
we introduce some new notations which we believe are easy to 
systematize. For an unpolarized nucleon target, one can introduce 
the unpolarized quark distribution $f_1(x, k_\perp)\equiv q(x,k_\perp)$ 
and time-reversal odd transversely-polarized quark distribution $h_1^\perp(x,k_\perp)
\equiv \delta q(x, k_\perp)$ arising from the final-state or 
initial-state interactions, where $x$ and $k_\perp$ are the longitudinal
momentum fraction and transverse momentum of the quark, respectively. 
For a longitudinally-polarized nucleon, one introduces a 
longitudinally-polarized quark distribution $g_{1L}(x, k_\perp) 
\equiv \Delta q_L(x, k_\perp)$ and a transversely-polarized distribution 
$h_{1T}^\perp(x, k_\perp) \equiv 
\delta q_L(x,k_\perp)$. Finally, for a transversely-polarized nucleon, 
one introduces a spin-independent distribution $f^\perp_{1T}(x, k_\perp)
\equiv q_T(x, k_\perp)$ arising from final and/or initial state interactions, 
and a longitudinally-polarized quark polarization 
$g_{1T}(x, k_\perp)=\Delta q_T(x,k_\perp)$, a symmetrical transversely-polarized 
quark distribution $h_{1T}(x, k_\perp)=\delta q_T(x, k_\perp)$ and an 
asymmetric transversely-polarized quark distribution
$h_{1T}^\perp(x, k_\perp)=\delta q_{T'}(x, k_\perp)$.

Out of the eight distributions, four of them are related to the nucleon
helicity flip and are the main interest in this subsection: unpolarized and 
longitudinally-polarized quark distributions
in a transversely-polarized target, $q_T(x, k_\perp)$ and $\Delta q_T(x, k_\perp)$,
and transversely-polarized quark distributions in an unpolarized 
and longitudinally-polarized nucleon, $\delta q(x, k_\perp)$ and $\delta q_L(x, k_\perp)$. 
Two of them, $q_T(x,k_\perp)$ and $\delta q(x, k_\perp)$, are non-zero only when
there are initial and final state interactions.

Unlike the usual Feynman parton distributions, there are some important subtleties 
about the transverse-momentum dependent parton distributions. First is  
gauge invariance and universality. Each of these distributions can be gauge-invariantized
in different ways depending on choices of gauge links and the results are different. 
Which one appears in a particular process requires a careful study. In the Drell-Yan process 
studied by Collins and Soper \cite{Collins:1981uk}, the distributions are 
defined in axial gauges. In deep-inelastic scattering, it has been shown 
that the gauge links are in the future direction along the light-cone. 
If the light-cone gauge is used, an additional gauge link at spatial 
infinity is required to render the distributions independent of additional gauge fixing 
\cite{Belitsky:2002sm}. Different versions of the distributions appear in different
hard processes imply that the universality of these distributions are lost.  
For example, the transverse-momentum dependent parton distributions measured
in DIS cannot be used to make predictions of the cross sections for Drell-Yan. 
For processes such as jet production in hadron-hadron scattering, it is not yet 
clear which version of the distributions is relevant. 

Another issue is infrared divergences. The distributions involving light-cone
gauge links have infrared divergences, which are cancelled after integrating out
parton transverse momentum. Physically, the infrared divergences can be understood
in the following way. The transverse-momentum dependent distributions can be measured
in DIS through single-jet production. However, the single-jet cross section is
not well-defined because the vertex correction has infrared divergence. 
The divergence is cancelled by the soft gluon radiation. Since a hard gluon radiation
is considered as a two-jet event, one must introduce an infrared cut-off to 
separate the single from double jet event. 
In the following discussion, we implicitly assume the
infrared cut-off is there when needed.

We now use the wave function of the previous section to calculate the 
helicity-flip parton distributions. We first consider the case of a transversely
polarized target. Introduce the quark density matrix, 
\begin{equation}
         {\cal M} = p^+\int \frac{d\xi^-d^2 \xi_\perp}{(2\pi)^3}
                e^{i(k^+\xi^--\vec{k}_\perp\cdot \vec{\xi}_\perp)}
            \langle PS| \overline \psi(0)L^\dagger_0 L_\xi \psi(\xi) |PS\rangle \ , 
\end{equation}
where $L_\xi$ is the gauge link along the light-cone in the covariant
gauges, $S^\mu$ is the polarization vector of the nucleon
normalized to $S_\mu S^\mu=-1$, $p^\mu$ is a light-cone vector such that 
$p^-=0$. The distributions we are interested can be obtained from ${\cal M}$ through
the expansion \cite{Boer:1998nt},
\begin{equation}
           {\cal M} = \frac{1}{2M} \left[q_T(x,k_\perp)\epsilon^{\mu\nu\alpha\beta}\gamma_\mu p_\nu k_\alpha S_\beta
          +  \Delta q_T(x, k_\perp) \gamma_5\not\! p (\vec{k}_\perp\cdot \vec{S}_\perp)
            + ... \right] \ , 
\end{equation}
where $M$ is the nucleon mass.
Inverting the above equation, we obtain
\begin{eqnarray}
           q_T(x, k_\perp)
           &=& -\frac{M}{2\epsilon^{ij}k_iS_j}
               \int \frac{d\xi^-d^2\xi_\perp}{(2\pi)^3}
             e^{i(k^+\xi^--\vec{k}_\perp\cdot \vec{\xi}_\perp)}
              \langle PS_\perp|\overline{\psi}(0)L_0^\dagger \gamma^+
                 L_\xi \psi(\xi)|PS_\perp\rangle \ ,  \nonumber \\
    \Delta q_T(x, k_\perp)
           &=& \frac{M}{2k_iS_i}
               \int \frac{d\xi^-d^2\xi_\perp}{(2\pi)^3}
 e^{i(k^+\xi^--\vec{k}_\perp\cdot \vec{\xi}_\perp)}
              \langle PS_\perp|\overline{\psi}(0)L_0^\dagger \gamma^+\gamma_5
                 L_\xi \psi(\xi)|PS_\perp\rangle \ . 
\end{eqnarray}
where the transversely-polarized nucleon in the $x$ direction
is related to the helicity states
by $|S_x\rangle = (|\uparrow\rangle + |\downarrow\rangle)/\sqrt{2}$. 

If these distributions are calculated in the light-cone gauge 
with the advanced boundary condition for the gauge potential, 
the gauge links can be ignored, but the wave function amplitudes
are complex \cite{Belitsky:2002sm}. Express the quark fields in 
term of Fock operators $(x>0)$,
\begin{eqnarray}
           q_T(x, k_\perp)
           &=& -\frac{M}{2\epsilon^{ij}k_iS_j}
               \frac{1}{4x (2\pi)^3 V_3}
              \langle PS_\perp|\sum_\lambda q^\dagger_\lambda(k) 
                q_\lambda(k)|PS_\perp\rangle \ ,  \nonumber \\
    \Delta q_T(x, k_\perp)
           &=& \frac{M}{2k_iS_i}
            \frac{1}{4x (2\pi)^3 V_3}
              \langle PS_\perp|\sum_\lambda (-1)^{(1/2-\lambda)}q^\dagger_\lambda(k) 
                q_\lambda(k)|PS_\perp\rangle \ , 
\end{eqnarray}
where $V_3$ is the 3-space volume. 
Using the above and the proton wave function from the previous section, 
we obtain
\begin{eqnarray}
      \Delta q_T(x,k_\perp)  &=& \frac{M}{k_\perp^2}
	\int d[1]d[2]d[3]\sqrt{x_12x_22x_3} ~{\rm Re}[F_q] \ , \nonumber \\ 
      q_T(x,k_\perp) &=& \frac{M}{k_\perp^2}\int d[1]d[2]d[3]  
	\sqrt{x_12x_22x_3} ~ {\rm Im}[F_q] \ . 
\end{eqnarray}
The functions $F_q$ for the u-quark is
\begin{eqnarray}
F_u&=&2\left\{ 
\delta^{(3)}(k-k_1)\tilde\psi^{(1,2)*}_j(1,2,3)\tilde\psi^{(3,4)}_j(1,2,3)-
\delta^{(3)}(k-k_2)\tilde\psi^{(1,2)}_j(1,2,3)\tilde\psi^{(3,4)*}_j(2,1,3)
\right.\nonumber\\
	&&+(\delta^{(3)}(k-k_1)+\delta^{(3)}(k-k_3))\tilde\psi^{(1,2)*}_j(1,2,3)(\tilde\psi^{(3,4)}_j(2,1,3)+\tilde\psi^{(3,4)}_j(2,3,1))
\nonumber\\&&
	+(\delta^{(3)}(k-k_1)+\delta^{(3)}(k-k_2))\tilde\psi^{(5-)*}(1,2,3)
\tilde\psi^{(6+)}(1,2,3)
\left. \right\} \ ,
\end{eqnarray}
where 
\begin{eqnarray}
\delta^{(3)}(k-k_i)&=&\delta(x-x_i)\delta^{(2)}(\vec{k}_\perp-\vec{k}_{i\perp}) \ ,\nonumber\\
\tilde\psi^{(1,2)}_j(1,2,3)&=&
\tilde\psi^{(1)}(1,2,3)k^j-
(k_1^xk_2^y-k_1^yk_2^x)\tilde\psi^{(2)}(1,2,3))k^i\epsilon^{ij}\nonumber \ ,\\
\tilde\psi^{(3,4)j}(1,2,3)&=&k_1^j\tilde\psi^{(3)}(1,2,3)+
	k_2^j\tilde\psi^{(4)}(1,2,3) \ ,\nonumber\\
\tilde\psi^{(5-)}(1,2,3)&=&\tilde\psi^{(5)}(1,2,3)-\tilde\psi^{(5)}(1,3,2)) \ ,\nonumber\\
\tilde\psi^{(6+)}(1,2,3)&=& 
-\vec{k}_{1\perp}^2\vec{k}_{3\perp}\cdot \vec{k}_\perp\tilde\psi^{(6)}(1,2,3)
	+(\vec{k}_{2\perp}^2\vec{k}_{3\perp}\cdot \vec{k}_\perp+
\vec{k}_{3\perp}^2\vec{k}_{2\perp}\cdot \vec{k}_\perp)\tilde\psi^{(6)}(2,1,3)
\nonumber\\&&
+\vec{k}_{1\perp}^2\vec{k}_2\cdot \vec{k}_\perp(\tilde\psi^{(6)}(1,3,2)
+\tilde\psi^{(6)}(2,3,1)) \ .
\end{eqnarray}
For the d-quark, on the other hand, we obtain
\begin{eqnarray}
F_d&=&2\left\{\delta^{(3)}(k-k_3)\tilde\psi^{(1,2)*}_j(1,2,3)\tilde\psi^{(3,4)}_j(1,2,3)\right.\nonumber\\&&-
\delta^{(3)}(k-k_2)\tilde\psi^{(1,2)}_j(1,2,3)(\tilde\psi^{(3,4)*}_j(2,1,3)+\tilde\psi^{(3,4)*}_j(2,3,1))
\nonumber\\&&
+\delta^{(3)}(k-k_3)\tilde\psi^{(5-)*}(1,2,3)\tilde\psi^{(6+)}(1,2,3)
 \left.\right\} \ .
\end{eqnarray}
If all the wave function amplitudes are real, $q_T(x,k_\perp)$ vanishes
identically. $q_T(x,k_\perp)$ was first introduce by Sivers
\cite{Sivers:1990cc,Sivers:1991fh} and it characterizes the azimuthal asymmetry in 
the quark transverse momentum distribution when the nucleon is transversely polarized.
There is much interest in the this distribution in interpreting the
single spin asymmetry measured in electron-proton deep inelastic scattering
\cite{Airapetian:1999ib,Airapetian:2001yk,Brodsky:2002cx,
Anselmino:1997qx,Ma:2001ie,Bacchetta:2002tk,Efremov:2001cz}.

The other two helicity-flip distributions can be obtained 
from the quark density matrix $M$ through the following projection
\cite{Boer:1998nt}
\begin{equation}
        {\cal M} = \frac{1}{2M}\left[\delta q(x,k_\perp)
              \sigma^{\mu\nu} k_\mu p_\nu
         + \delta q_L(x, k_\perp)
           i\sigma^{\mu\nu}\gamma_5 p_\mu k_\nu (S\cdot n)+...\right] \ . 
\end{equation}
The transversely-polarized quark distribution in an unpolarized
nucleon $\delta q(x, k_\perp)$ is novel, and has the similar 
physical origin as that of 
the experimental phenomenon where a hyperon produced in
unpolarized elastic proton-proton scattering is polarized 
in the direction perpendicular to the production plane 
\cite{Bunce:1976yb,Liang:2000gz}.
Invert the above equation, we arrive at
\begin{eqnarray}
           \delta q (x, k_\perp)
           &=& -\frac{M}{2k_i}
               \int \frac{d\xi^-d^2\xi_\perp}{(2\pi)^3}
  e^{i(k^+\xi^--\vec{k}_\perp\cdot \vec{\xi}_\perp)}
              \langle P|\overline{\psi}(0)L_0^\dagger \sigma^{+i}
                 L_\xi \psi(\xi)|P\rangle \ , \nonumber \\
    \delta q_L(x, k_\perp)
           &=& \frac{M}{2\epsilon^{ij}k^j}
               \int \frac{d\xi^-d^2\xi_\perp}{(2\pi)^3}
  e^{i(k^+\xi^--\vec{k}_\perp\cdot \vec{\xi}_\perp)}
              \langle PS_{||}|\overline{\psi}(0)L_0^\dagger \sigma^{+i}
                 L_\xi \psi(\xi)|PS_{||}\rangle \ .  
\end{eqnarray}
where $i$ can take either $x$ or $y$. 

Inserting the plane wave expansion for the quark fields, we get
\begin{eqnarray}
           \delta q(x, k_\perp)
           &=& -\frac{M}{k^x}
               \frac{1}{4x (2\pi)^3 V_3}
              i \langle P|q^\dagger_{\downarrow}(k) q_{\uparrow}(k)
           - q^\dagger_{\uparrow}(k) q_{\downarrow}(k)|P\rangle \ ,  \nonumber \\
    \delta q_L(x, k_\perp)
           &=& \frac{M}{\epsilon^{xj}k^j}
             \frac{1}{4x (2\pi)^3 V_3}
              i\langle PS_{||}|q^\dagger_{\downarrow}(k) q_{\uparrow}(k)
           - q^\dagger_{\uparrow}(k) q_{\downarrow}(k)
   |PS_{||}\rangle  \ . 
\end{eqnarray}
Using the above and the proton wave function from the previous section,
we obtain,
\begin{eqnarray}
      \delta q_L(x,k_\perp)  &=& \frac{M}{k_\perp^2}\int d[1]d[2]d[3]\sqrt{x_12x_22x_3} 
	~{\rm Re}[H_q] \ , \nonumber  \\ 
      \delta q(x,k_\perp) &=& \frac{M}{k_\perp^2}\int d[1]d[2]d[3] \sqrt{x_12x_22x_3}
	~{\rm Im}[H_q] \ . 
\end{eqnarray}
The function $H_q$ is 
\begin{eqnarray}
     H_u &=& 2
\left\{-\delta^{(3)}(k-k_{1})\tilde\psi^{(1,2)'}_j(1,2,3)(
\tilde\psi^{(3,4)*}_j(3,1,2)+\tilde\psi^{(3,4)*}_j(3,2,1))
\nonumber \right.\\
&&-\delta^{(3)}(k-k_{2})(\tilde\psi^{(1,2)'}_j(1,3,2)+
\tilde\psi^{(1,2)'}_j(2,3,1))\tilde\psi^{(3,4)*}_j(1,2,3)\nonumber\\&&
-\delta^{(3)}(k-k_{2})\tilde\psi^{(1,2)*}_j(1,2,3)
(k_2^j \tilde\psi^{(5-)}(2,1,3)+
	k_{1}^j\tilde\psi^{(5-)}(1,2,3))\nonumber \\&&+
\delta^{(3)}(k-k_{1})(\tilde\psi^{(6+)*}(1,2,3)\tilde\psi^{(3)}(1,2,3)+
\tilde\psi^{(6+)*}(2,1,3)\tilde\psi^{(4)}(1,2,3))
\left. \right\},
\end{eqnarray}
where $\tilde\psi^{(1,2)'}_j(1,2,3)=
\tilde\psi^{(1)}(1,2,3)k^j+
(k_1^xk_2^y-k_1^yk_2^x)\tilde\psi^{(2)}(1,2,3))k^i\epsilon^{ij}$. 
For the down quark,
\begin{eqnarray}
     H_d &=&  2
\left\{\delta^{(3)}(k-k_{3})\tilde\psi^{(1,2)'}_j(1,2,3)
\tilde\psi^{(3,4)*}_j(1,2,3)\nonumber \right.\\
&&-\delta^{(3)}(k-k_{2})\tilde\psi^{(1,2)*}_j(1,2,3)
(k_1^j \tilde\psi^{(5-)}(1,2,3)+k_3^j \tilde\psi^{(5-)}(3,2,1))\nonumber\\
&&+\delta^{(3)}(k-k_{1})(\tilde\psi^{(6+')*}(1,2,3)\tilde\psi^{(3)}(1,2,3)-
\tilde\psi^{(6+)*}(2,3,1)\tilde\psi^{(4)}(1,2,3))
\left. \right\} \ , 
\end{eqnarray}
where 
\begin{eqnarray}
\tilde\psi^{(6+')}(1,2,3)&=& 
	(\vec{k}_{2\perp}^2\vec{k}_{3\perp}\cdot \vec{k}_\perp+
\vec{k}_{3\perp}^2\vec{k}_{2\perp}\cdot \vec{k}_\perp)
(\tilde\psi^{(6)}(2,1,3)+\tilde\psi^{(6)}(3,1,2))
\nonumber\\&&
+\vec{k}_{1\perp}^2\vec{k}_2\cdot \vec{k}_\perp\tilde\psi^{(6)}(2,3,1)
+\vec{k}_{1\perp}^2\vec{k}_3\cdot \vec{k}_\perp\tilde\psi^{(6)}(3,2,1) \ .
\end{eqnarray}
If all amplitudes are real, $\delta q(x,k_\perp)$ vanishes identically.

\subsection{Twist-Three Parton Distributions}

It has been known for many years that the helicity flip
phenomena in inclusive scattering are of higher twist
effects (twist-three, to be exact). For example, the single
spin asymmetry measured in pion production in polarized baryon
(proton, hyperons) and unpolarized target scattering arises
from hadron helicity flip \cite{Barone:2001sp}. The asymmetry is expected to 
vanish asymptotically like $1/k_\perp$, where $k_\perp$
is the pion transverse momentum \cite{Qiu:1991pp}.   
The other example is the $g_2$ structure function 
measured in inclusive DIS with transversely polarized nucleon targets. 
The structure function arises from the interference between 
photon scatterings with longitudinal and transverse polarizations 
and involves the helicity flip of the nucleon
target because of the angular momentum conservation 
\cite{Jaffe:1992ra}. The associated asymmetry goes like $1/Q$ where
$Q$ is the virtual photon mass. 

In both examples, the baryon helicity flip must be reflected
in the hard scattering subprocesses. However, if 
the quark transverse momentum is neglected as one usually does in the 
Feynman parton model, the helicity flip cannot be managed through a 
massless quark; this has been considered as one of the
difficulties in perturbative QCD \cite{Kane:1978nd,Jackson:1989ph}.
It turns out that the leading contribution to the baryon helicity
flip comes from the quark orbital angular momentum and associated
transversely-polarized gluons \cite{Efremov:1985ip,Qiu:1991pp}. 
This can be seen from the following light-cone expression
for the $g_2(x)$ structure function \cite{Jaffe:1992ra}
\begin{equation}
    g_T(x) = \frac{-1}{4xM}
\int \frac{d\lambda}{2\pi}
   e^{i\lambda x}
       \langle PS_\perp|\overline{\psi}(0)
          \not\! n\gamma_5 \not\!S_\perp i\not\! D_\perp
        (\lambda n) \psi(\lambda n)|PS_\perp\rangle + {\rm h. c.} \ .     
\end{equation}
In our approximation, the gluon potential in the
covariant derivative is neglected. Using the
three-quark wave function, we find,
\begin{equation}
  g_T(x) = \frac{1}{2xM^2}\int  k_\perp^2 \Delta 
   q_T(x, k_\perp) d^2\vec{k}_\perp \ . 
\end{equation}
In fact, the above result holds for any nucleon
wave function so long as the
gluon potential in the covariant derivative 
can be ignored \cite{Mulders:1996dh}. 
A word of caution, however, is appropriate. When the gauge
potential is neglected, $g_T(x)$ is no longer gauge invariant. 
How it is possible then to express $g_T(x)$ in terms of the
the gauge-invariant $\Delta q_T(x, k_\perp)$? The answer is that
the above relation is only true in light-cone gauge with the
advanced prescription for the light-cone singularity. 
In any other gauge, the two have no relation.

It is interesting to note that the rotational invariance
imposes the following condition on $g_T(x)$ 
\cite{Feynman:1972}
\begin{equation}
          \int^1_0 dx g_T(x) = \int^1_0 dx g_1(x) \ . 
\end{equation}
Therefore, when the gluons are neglected,
the rotational symmetry demands
the quarks have non-vanishing orbital angular 
momentum. This is indeed true for a massless quark 
in the MIT bag model. However, it is possible that $g_T(x)$
may have a delta function at $x=0$ \cite{Burkardt:2001iy}. 
In this case, the above is useless in constraining 
$g_T(x)$ at any non-zero $x$. 

Similar analysis can be carried out for the twist-three 
distribution $h_L(x)$ which has the following expression,
\begin{equation}
    h_L(x) = \frac{-1}{4xM}
\int \frac{d\lambda}{2\pi}
   e^{i\lambda x}
       \langle PS_\perp|\overline{\psi}(0)
          \not\! n\gamma_5  i\not\! D_\perp
        (\lambda n) \psi(\lambda n)|PS_\perp\rangle + {\rm  h.c.}\ .     
\end{equation}
Using the proton wave function, we find
\begin{equation}
       h_L(x) = -\frac{1}{xM^2} \int {k_\perp^2} 
\delta q_L(x, k_\perp) d^2k_\perp  \ , 
\end{equation}
which is again consistent with \cite{Mulders:1996dh}
apart from a factor of 2. 
The twist-three distribution $h_L(x)$ 
can be measured in Drell-Yan scattering with
the transversely-polarized protons scattering on 
longitudinally-polarized protons \cite{Jaffe:1992ra}. 

\subsection{Generalized Parton Distributions}

Generalized parton distributions (GPDs) were originally introduced to describe
the angular momentum structure of the nucleon \cite{Ji:1997ek,Ji:1997nm}.
D. Mueller et al. encountered the same distributions earlier in 
searching for an observable whose scale evolution interpolates between
those of the Feynman parton distributions and the quark distribution 
amplitudes in mesons \cite{Muller:1994fv}. Radyushkin 
defined the so-called double distributions which contain essentially
the same information \cite{Radyushkin:1997ki}. The new distributions 
contain all the form factors of the twist-two operators \cite{Ji:1998pc} 
and, in a special kinematic limit, reduce to the Feynman distributions. 
In this subsection, we calculate the helicity-flip GPDs using the three-quark
wave function of the proton. 

At the leading twist, there are eight GPDs for each quark flavor
\cite{Hoodbhoy:1998vm,Diehl:2001pm}, out of which four involve 
the nucleon helicity flip. In this paper we consider
only $E(x, \xi, Q^2)$ which is directly relevant for the 
spin structure of the nucleon \cite{Hoodbhoy:1998yb}. Extending the 
calculation to other three is straightforward and will not be presented
here. The definition of $E(x, \xi, Q^2)$ follows from \cite{Ji:1997nm}
\begin{eqnarray}
    && \frac{1}{2}\int \frac{d\lambda}{ 2\pi} e^{i\lambda x}
    \left\langle P'\left|\overline \psi_q \left(-\frac{\lambda}{2}n\right)
      \not \! n {\cal P}e^{-ig\int^{-\lambda/2}_{\lambda/ 2}
       d\alpha ~n\cdot A(\alpha n)} 
    \psi_q\left(\frac{\lambda }{ 2}n\right) \right| P\right\rangle 
    \nonumber \\
  && = H_q(x, \xi, Q^2)~ \frac{1}{2}\overline U(P')\not\! n U(P)
    + E_q(x, \xi, Q^2)~ \frac{1}{2}\overline U(P') \frac{i\sigma^{\mu\nu}
  n_\mu \Delta_\nu }{ 2M} U(P) \ ,
\label{string}
\end{eqnarray}  
where the explanation of various notations can be found in the
original references.

To evaluate $E(x, \xi, Q^2)$, we choose a coordinate system in 
which the momentum transfer $q^\mu$ vanishes along the $+$ direction,
\begin{equation}
     q^+=0,~~ Q^2=-q^2_\perp \ . 
\end{equation}
Then the initial and final nucleons have the same 
$P^+$ momenta. Their transverse momenta are equal in magnitude 
and opposite in direction (a Breit frame in the transverse direction). 
The energy-momentum conservation constrains $q^-=0$. 
With these choices, we can only calculate the distribution at the skewness 
variable $\xi=0$. [For $\xi\ne 0$, one can use the coordinates 
in Ref. \cite{Brodsky:2000xy}.]

Our calculation follows a similar formula for $F_2(Q^2)$ 
in Ref. \cite{Brodsky:1980zm}, 
\begin{equation}
       -(q^x-iq^y) E(x,\xi=0,Q^2) \frac{P^+}{M}
   = \int \frac{d\lambda}{2\pi} e^{i\lambda x}
    \left\langle P'\left|\overline \psi \left(-\frac{\lambda }{2}n\right)
      \gamma^+  \psi\left(\frac{\lambda }{ 2}n\right) \right| P\right\rangle \ . 
\end{equation}
For $x>0$, the bilocal operator reduces to quark creation and annihilation
operators in the momentum space, and for $x<0$, it becomes antiquark
creation and annihilation. The later contribution
vanishes in the valence approximation.

Using the proton wave function from the previous section, one finds
\begin{eqnarray}
E_u(x,\xi=0, Q^2)&=&\frac{2M}{-q^x+iq^y}\int d[1]d[2]d[3] \sqrt{x_12x_22x_3}
\left\{A^{(1,2)}\phi^{(3,4)}(1,2,3)\right.\nonumber\\
&&-A^{(3,4)}\phi^{(1,2)'}(1,2,3)-
	A^{(5)}\phi^{(6)}(1,2,3)+A^{(6)}\phi^{(5)}(1,2,3)
\left.\right\}  \ , 
\end{eqnarray}
where
\begin{eqnarray}
\phi^{(1,2)}(1,2,3)&=&\tilde\psi^{(1)}(1,2,3)+
i(k_1^xk_2^y-k_1^yk_2^x)\tilde\psi^{(2)}(1,2,3) \ ,\nonumber\\
\phi^{(1,2)'}(1,2,3)&=&\tilde\psi^{(1)}(1,2,3)-
i(k_1^xk_2^y-k_1^yk_2^x)\tilde\psi^{(2)}(1,2,3) \ ,\nonumber\\
\phi^{(3,4)}(1,2,3)&=&k_1^-\tilde\psi^{(3)}(1,2,3)
	+k_2^-\tilde\psi^{(4)}(1,2,3)\ ,\nonumber\\
\phi^{(5)}(1,2,3)&=&k_1^+(\tilde\psi^{(5)}(1,2,3)
	-\tilde\psi^{(5)}(1,3,2))\ ,\nonumber\\
\phi^{(6)}(1,2,3)&=&k_1^-k_3^-\tilde\psi^{(6)}(1,2,3)
		-k_1^-k_2^-\tilde\psi^{(6)}(1,3,2) \ . 
\end{eqnarray}
And the functions $A$ are defined through
\begin{eqnarray}
A^{(1,2)}&=&\delta(x-x_1)\phi^{(1,2)*}(2',1'',3')+
\delta(x-x_2)(2\phi^{(1,2)*}(2'',1',3')+\phi^{(1,2)*}(3',1',2''))\nonumber\\
&&+\delta(x-x_3)(\phi^{(1,2)*}(2',1',3'')+\tilde\psi^{(1,2)*}(3'',1',2'))\ ,
\nonumber\\
A^{(3,4)}&=&\delta(x-x_2)\phi^{(3,4)*}(2'',1',3')+
\delta(x-x_1)(2\phi^{(3,4)*}(2',1'',3')+\phi^{(3,4)*}(2',3',1''))\nonumber\\
&&+\delta(x-x_3)(\phi^{(3,4)*}(2',1',3'')+
	 	\psi^{(3,4)*}(2',3'',1'))\ ,\nonumber\\
A^{(5)}&=&\delta(x-x_1)(\phi^{(5)*}(1'',2',3')+
		\phi^{(5)*}(2',1'',3'))\nonumber\\
&&+\delta(x-x_2)(\phi^{(5)*}(1',2'',3')+
		\phi^{(5)*}(2'',1',3'))\ ,\nonumber\\
A^{(6)}&=&\delta(x-x_1)(\phi^{(6)*}(1'',2',3')+
	\phi^{(6)*}(2',1'',3'))\nonumber\\
&&+\delta(x-x_2)(\phi^{(6)*}(1',2'',3')+
	\phi^{(6)*}(2'',1',3')) \ , 
\end{eqnarray}
where the transverse coordinates are 
$i'=\vec{k}_i-x_i\vec{q}_\perp$ and $i''=\vec{k}_i+(1-x_i)\vec{q}_\perp$,  
$k_i^\pm=k_i^x\pm i k_i^y$. The result
depends on the interference of the wave function amplitudes with
different quark orbital angular momentum.

Similarly for the d-quark, 
\begin{eqnarray}
E_d(x,\xi=0,Q^2)&=&\frac{2M}{-q^x+iq^y}\int d[1]d[2]d[3] \sqrt{x_12x_22x_3}
\left\{B^{(1,2)}\phi^{(3,4)}(1,2,3)\right.\nonumber\\
&&-B^{(3,4)}\phi^{(1,2)'}(1,2,3)-
	B^{(5)}\phi^{(6)}(1,2,3)+B^{(6)}\phi^{(5)}(1,2,3)
\left.\right\} \ , 
\end{eqnarray}
where
\begin{eqnarray}
B^{(1,2)}&=&\delta(x-x_3)\phi^{(1,2)*}(2',1',3'')+
\delta(x-x_1)(\phi^{(1,2)*}(2',1'',3')+\phi^{(1,2)*}(3',1'',2'))\ ,
\nonumber\\
B^{(3,4)}&=&\delta(x-x_3)\phi^{(3,4)*}(2',1',3'')+\delta(x-x_2)(\phi^{(3,4)*}(2'',1',3')+\phi^{(3,4)*}(2'',3',1'))
\ ,\nonumber\\
B^{(5)}&=&\delta(x-x_3)(\phi^{(5)*}(1',2',3'')+
		\phi^{(5)*}(2',1',3''))\ ,\nonumber\\
B^{(6)}&=&\delta(x-x_3)(\phi^{(6)*}(1',2',3'')+
	\phi^{(6)*}(2',1',3''))\ .
\end{eqnarray}
Although we have allowed $\vec{q}$ to be arbitrary in the above formulas, 
it is simpler to take it in the
$x$ direction. With this choice, it is not difficult
to see that the distributions are real. 

\subsection{The Pauli Form Factor}

The $F_2$ form factor can be obtained from the sum rule \cite{Ji:1997ek},
\begin{equation}
           F_2(Q^2) = \int dx [e_uE_u(x, \xi, Q^2)+e_dE_d(x,\xi, Q^2)] \ , 
\end{equation}
where we have neglected the strange quark contribution. 
Recently, the Pauli form factor of the proton has been extracted
from the recoil polarization \cite{Jones:1999rz,Gayou:2001qd}, 
which differs significantly from the previous extraction. 
In the valence approximation, independent of $Q^2$,
a non-zero $F_2(Q^2)$ is an indication of non-zero 
orbital angular momentum of the quarks in light-cone 
quantization. This point has been stressed in a number 
of recent papers which seek to explain the new Jlab data
\cite{Miller:2002qb,Ralston:2002ep}.  

In large-$Q^2$ limit, we can use the asymptotic behavior
of amplitudes to derive the asymptotic behavior of $F_2(Q^2)$. 
The result will be related to the leading and twist-three 
light-cone amplitudes introduced by Braun et al. \cite{Braun:2000kw}. 
We leave this subject to a future publication. 

Finally, we pointed out that the $B$-form factor of the
gravitational current can be obtained from the second moment
of $E(x, \xi, Q^2)$ \cite{Ji:1997ek} and is also helicity-flipping. In Ref. 
\cite{Brodsky:2000ii}, the $B$ form factor is calculated
for the electron in quantum electrodynamics, with explicit 
involvement of orbital angular momentum.

\section{Summary and Conclusion}

This paper is motivated by determining the 
components of the proton wave function 
with non-zero orbital angular momentum. This 
of course can only be down in a certain truncation scheme, because
the field-theoretical nucleon has an infinite number of 
non-vanishing amplitudes and there is no finite set of 
experiments which can determine them completely. 

We use the valence quark models as a guide to 
truncate the light-cone expansion of the proton 
wave function. We can view this either as a starting 
point for a systematic expansion or as an effective
theory in which gluons and sea quarks have been 
``integrated out". 

We classify the light-cone amplitudes with three quarks,
and find that six independent amplitudes are needed for
a complete description: two with $L_z=0$,
three with $L_z=1$, and one with $L_z=2$. If the light-cone 
gauge fixing is invariant under combined parity and time-reversal,
these amplitudes are real. Otherwise, they are complex.
Apart from one of the $L_z=0$ amplitudes,
the other five contain nontrivial quark orbital motion. 

We calculated a number of nuclear observables which depend
at least linearly on the amplitudes with non-zero orbital angular 
momentum. Therefore, a non-vanishing result of the experimental 
measurement is an unambiguous indication of the quark orbital 
angular momentum, barring the neglected gluonic contributions 
mentioned above.

It is our hope that multiple observables of this type can be
explained by a unified set of phenomenological amplitudes 
with nontrivial orbital angular momentum. If not, one may 
systematically go beyond the minimal Fock component by including,
for example, the
gluon contributions. Ultimately, we hope to have a
semi-realistic picture about the intrinsic deformation of 
the proton from experimental data.

The authors thank S. Brodsky for a number of useful discussions. 
This work was supported by the U. S. Department of Energy via 
grants DE-FG02-93ER-40762.

\bibliography{protonref}

\end{document}